# Light-induced switching in the back-gated organic transistors with built-in conduction channel

V. Podzorov [*], V. M. Pudalov, and M. E. Gershenson

*Department of Physics and Astronomy, Rutgers University, Piscataway, New Jersey 08854*

(29 June 2004)

We report on observation of a light-induced switching of the conductance in the back-gated organic field-effect transistors (OFETs) with built-in conduction channel. In the studied devices, the built-in channel is formed owing to the self-sensitized photo-oxidation of rubrene surface. In the dark, the back gate controls the charge injection from metal contacts into the built-in channel: the high-current *ON* state corresponds to zero or negative back-gate voltage; the low-current *OFF* state – to a positive back-gate voltage that blocks the Schottky contacts. Illumination of the OFET in the *OFF* state with a short pulse of light switches the device into the *ON* state that persists in the dark for days. The *OFF* state can be restored by cycling the back gate voltage. The observed effect can be explained by screening of the back-gate electric field by the charges photo-generated in the bulk of organic semiconductor.

The organic field effect transistors (OFETs) represent one of the key components in the organic electronics [1]. For many applications, the control of transistor operation with light would be desirable. However, until recently, the light-induced effects in OFETs were limited to observation of a photoconductivity response (see, e.g., [2]). One of the problems that hinder the study of the photo-induced processes in conventional thin-film transistors is a large concentration of structural defects in organic thin films that trap light-generated carriers and act as the recombination centers. Fabrication of the single-crystal OFETs [3,4,5,6,7], in which concentration of defects is significantly reduced, opens new opportunities for studies of light-induced effects in organic semiconductor devices.

In this paper, we report on a novel light-induced effect in organic transistors: under illumination by a short pulse of light, the transistor switches from the low-current *OFF* state into the high-current *ON* state, which persists in the dark for days. This effect might be potentially interesting for light sensors and light-addressable memory applications.

We observed this effect in an unconventional *back-gated OFET with a built-in conduction channel*. The sketch of the back-gated OFET based on a single crystal of rubrene is shown in Fig. 1a. The OFETs have been fabricated using the parylene gate-dielectric technique that we used for fabrication of high-mobility OFETs based on free-standing organic molecular crystals [3]. The single crystals of rubrene have been grown from the vapor phase in a stream of ultra-pure hydrogen in a horizontal reactor [8]. Typically, thin flat crystals with dimensions of 1-3 mm in the (a,b) basal plane and 0.1-0.2 mm along the c-axis were used. The source/drain contacts and the gate electrode were formed on the opposite (a,b)-facets of the crystal, either by thermal evaporation of silver through a shadow mask or by painting with the colloidal graphite. These contacts are efficient injectors of *p*-type carriers in rubrene [3]. The injected carriers propagate along a thin built-in conduction channel, formed on the top surface of rubrene due to the *self-sensitized photo-oxidation* (see below). Other processes that may lead to formation of the built-in conduction channel include, for instance, charge transfer at the interface between two organic materials [9] and the dipole interaction between organic surfaces and self-assembled monolayers [10]. The thermally evaporated back gate is isolated from the crystal by a 0.5-μm-thick film of parylene and covers the entire back surface of the device. All the measurements described below have been performed at room temperature.

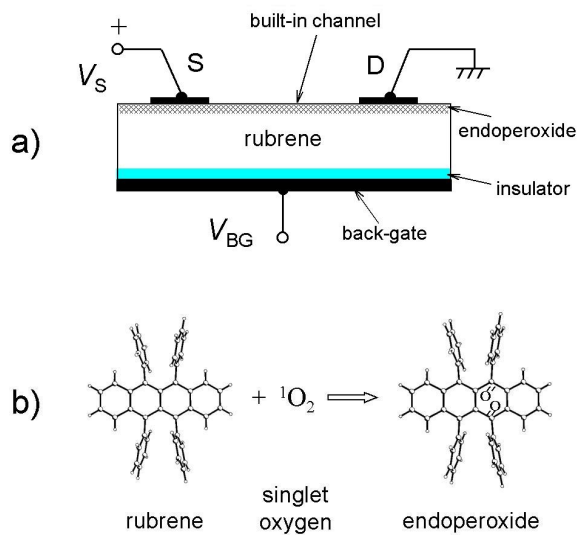

FIG. 1. (a) The rubrene back-gated OFET with the built-in conduction channel. The source/drain contacts and the conduction channel are on the top facet of the crystal; the back gate is isolated from the bottom facet with thin parylene film. (b) The reaction of the self-sensitized photo-oxidation of rubrene to endoperoxide by singlet molecular oxygen, responsible for the formation of the built-in surface conduction in rubrene [13,14].

Operation of the back-gated rubrene OFET is illustrated in Fig. 2. Application of a negative voltage to the back gate ($V_{BG}$), separated from the *p*-type channel by ~ 0.1-0.2 mm thick crystal, does not affect significantly the high-current *ON* state. However, application of a positive $V_{BG}$ in the dark reduces the source-drain current ($I_{SD}$) by many orders of magnitude by blocking the injection of carriers from the source contact into the conduction channel. The back gate controls the resistance of the Schottky contacts without directly affecting the band "bending" on the top surface of semiconductor, determined solely by the surface-restricted interactions. At a sufficiently large positive $V_{BG}$, the current through the device in the dark becomes very small (tens of fA) - much lower than the *OFF* current in the front-gated devices [3]. This is due to the fact that the back-gate electric field blocks the charge injection over *the entire area of the contacts*, whereas the front gate affects the injection only along the perimeter of the contacts. In the dark, the *OFF* state of our devices is stable: it lasts while positive $V_{BG}$ is maintained. The *ON* state can be restored by zeroing the back gate voltage.

---

[*] Electronic mail: podzorov@physics.rutgers.edu

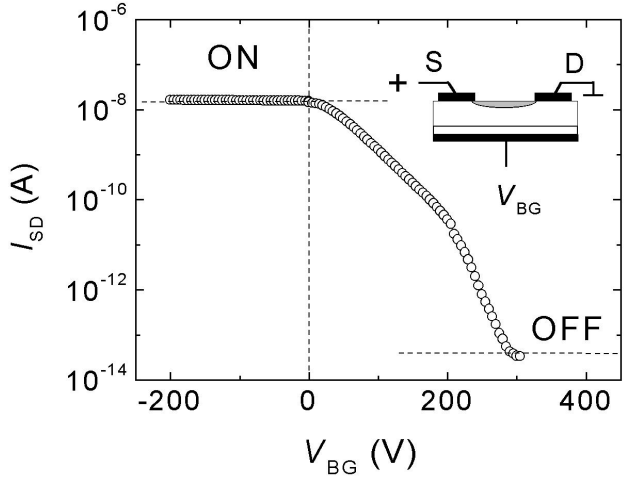

**FIG. 2.** The dependence of the source-drain current, $I_{SD}$, on the back-gate voltage, $V_{BG}$, measured at fixed $V_S = +10$ V in the dark. The sketch of the back-gated device is shown in the inset. The crystal thickness $h = 0.2$ mm, $L \sim W \sim 1$ mm, parylene thickness $d = 0.5$ μm $<< h$. At $V_{BG} > 300$ V, the current measurements are limited by the sensitivity of our experimental set-up ($\sim 10^{-14}$ A).

Despite the apparent similarity of the structure of these back-gated OFETs to the conventional "top-contact" organic thin-film transistors (OTFTs) [1], the operation principles for these devices are different. In the studied devices, the built-in conduction channel is formed at the semiconductor surface that is *opposite* to the gate [11]. The channel is separated from the gate not only by a layer of dielectric, but also by an optically active organic semiconductor – this feature of the back-gated OFETs is crucial for observation of the unusual optical response.

A relatively large conductivity of the device in the *ON* sate ($V_{BG} \leq 0$) is due to the built-in surface conduction channel [12]. The surface restricted oxidation of rubrene [13,14], responsible for this effect, occurs under illumination of as-grown crystals with visible light in air. Rubrene is oxidized by the singlet oxygen $^1O_2$ (the first excited state of the molecular oxygen, see Fig. 1b), which is produced in the process of photosensitization – the energy transfer from a photo-exited aromatic hydrocarbon with a long-lived triplet state (the sensitizer) to the molecular oxygen in the ground state ($^3O_2$) [14]. Rubrene is one of the most efficient organic sensitizers [15]. As the result of the *self-sensitized photo-oxidation*, a layer of *endoperoxide* grows at the rubrene surface exposed to air under illumination with visible light (Fig. 1b). Contrary to oxidation of disordered organic thin-films, oxidation of monolithic single crystals is restricted to a very thin surface layer [13]. A thin and uniform surface dipole layer created by the endoperoxide results in a violation of the flat-band condition at the surface and formation of a potential well filled with charge carriers injected from the contacts. Due to the built-in channel, the surface conductivity of as-grown free-standing crystals of rubrene in the basal (a,b)-plane is by a factor of $\sim 10^3$ - $10^6$ greater than that of the crystals of anthracene, tetracene, and other aromatic compounds that are not efficient sensitizers [13,15].

The photo-oxidation origin of the high conductivity of a free rubrene surface can be verified by measuring $I_{SD}$ at different ambient/illumination conditions. High conductivity of as-grown crystals can be reduced by many orders of magnitude due to de-sorption of oxygen by keeping the crystal at pressure $\sim 10^{-7}$ - $10^{-6}$ Torr in the dark for a few hours. Flashing the vacuum chamber with air does not significantly affect the reduced conductivity provided that the sample remains in the dark. However, a brief illumination of the sample with visible light in air restores the initial (high) value of σ, which persists in the dark.

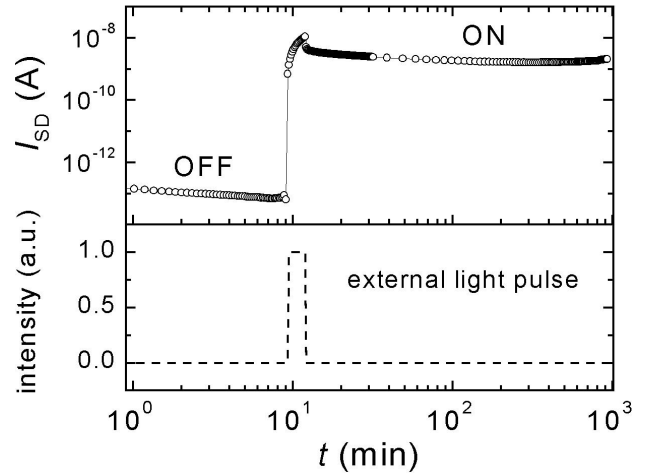

**FIG. 3.** The effect of a light pulse on the conductivity of the depleted back-gated rubrene OFETs. The source-drain current, $I_{SD}$, was measured at $V_S = +10$ V. The initial low-current *OFF* state in the dark was prepared by "blocking" the Schottky contacts by a large positive $V_{BG} = +300$ V, which was fixed for the whole duration of the experiment. Illumination of the front surface results in a factor-of-$10^3$ increase of $I_{SD}$ that persists after the light is off.

The back-gated OFETs with a built-in conduction channel demonstrate an intriguing photo-response. Illumination of the depleted devices in the *OFF* state with a short pulse of light leads to the switching into the *ON* state (Fig. 3) [16]. Interestingly, after the light pulse ends, the device remains in the *ON* state for days, despite the fact that the positive $V_{BG}$ is maintained. The *OFF* state can be restored by zeroing the back gate voltage and re-applying a positive $V_{BG}$ in the dark. The switching effect does not depend on the source-drain voltage ($V_S$): turning $V_S$ off and restoring it in either of the states does not alter the transistor state. The current in the *ON* state varies linearly with $V_S$.

We propose the following explanation for the observed effect. The *n*- and *p*-type carriers, photo-generated in the bulk of the organic crystal, are spatially separated in the electric field of the back gate. The *n*-type carriers migrate through the bulk toward the back gate (at $V_{BG} > 0$), whereas the *p*-type carriers are repelled toward the conducting channel and escape through the contacts. The *n*-type carriers, accumulated near the interface with the insulated gate, are unable to escape from the sample because the leakage current through the insulating layer of parylene is negligibly small. They form a negative space charge that screens the back gate electric field even after the light is off. Screening of the gate field "opens" the Schottky contacts, and the current starts flowing through the built-in channel on the top surface of the crystal. When the positive $V_{BG}$ is removed, the *n*-type carriers escape through the contacts, and repeated application of the positive $V_{BG}$ in the dark switches the device in the *OFF* state again. The response of the devices to light pulses, presumably limited by the time of diffusion of non-equilibrium charge carriers in the bulk, is fast owing to the high mobility of carriers in the single crystals of rubrene [3]. We did not observe any hysteresis in the device operation, which is consistent with a low density of charge traps in high-quality rubrene crystals.

Note that the observed phenomenon is different from a slow relaxation of the light-induced conductivity in the polymer FETs [2], which was attributed to the persistent photoconductivity. In the case of the studied rubrene FETs with a typical mobility of charge carriers ~10 cm$^2$/Vs, the mobile non-equilibrium carriers would be driven out of the conduction channel by a typical source-drain electric field ~ 100 V/cm within 10$^{-4}$ s.

To summarize, we fabricated and characterized the back-gated organic transistors with the built-in conduction channel formed due to a charge-transfer reaction at the organic surface. The back gate modulates the current in these devices by blocking the injection of *p*-type carriers from the contacts into the built-in channel. We have observed switching of these devices from the *OFF* state into the persistent *ON* state induced by a short pulse of light. The effect is attributed to the screening of the back-gate electric field by the non-equilibrium carriers photo-generated in the bulk and accumulated at the surface that faces the back gate electrode. The *OFF* state can be restored by cycling the back gate voltage in the dark. The absence of grain boundaries and well-ordered organic surface in the single-crystal OFETs facilitates observation and interpretation of this effect, as well as, more broadly, the study of surface restricted interactions and their effect on the electronic properties of organic electronic devices. The self-latching character of the observed effect opens an opportunity for using the organic field-effect devices with built-in conduction channel as the light-addressed memory elements and light sensors.

This work has been supported by the NSF grant DMR-0405208.